\documentclass[
aps, prl,
floatfix,
preprintnumbers,
reprint,
superscriptaddress,
amsmath, amssymb,
longbibliography,
]{revtex4-2}
\usepackage{float}
\usepackage{changepage}

\usepackage{soul}

\usepackage[utf8]{inputenc}
\usepackage{graphicx} 
\usepackage{bm}
\usepackage[breaklinks=true]{hyperref}
\hypersetup{bookmarksnumbered, pdfpagemode=UseOutlines, 
	pdfauthor={F.\ Feringa}, 
	pdftitle={Thermally generated Magnons in Quasi Two Dimensional Antiferromagnets}, 
	pdfdisplaydoctitle, 
	colorlinks=true, citecolor=blue, filecolor=blue, linkcolor=blue, urlcolor=blue}


\begin{document}
	
	\title{The spin-flop
		transition in the quasi two dimensional antiferromagnet MnPS$_{3}$ detected via thermally generated magnon transport}
	
	\author{F.\ \surname{Feringa}}
	\email[e-mail: ]{F.Feringa@rug.nl}
	\affiliation{Physics of Nanodevices, Zernike Institute for Advanced Materials, University of Groningen, 9747 AG Groningen, The Netherlands}
	\author{J.\ M.\ \surname{Vink}}
	\affiliation{Physics of Nanodevices, Zernike Institute for Advanced Materials, University of Groningen, 9747 AG Groningen, The Netherlands}
	\author{B.\ J.\ \surname{van Wees}}
	\email[e-mail: ]{B.J.van.Wees@rug.nl}
	\affiliation{Physics of Nanodevices, Zernike Institute for Advanced Materials, University of Groningen, 9747 AG Groningen, The Netherlands}

	\date{\today}

	\begin{abstract}
	 We present the detection of the spin-flop transition in the antiferromagnetic van der Waals material MnPS$_{3}$ via thermally generated nonlocal magnon transport using permalloy detector strips. The inverse anomalous spin Hall effect has the unique power to detect an out-of-plane spin accumulation \cite{Das2018a} which enables us to detect magnons with an out-of-plane spin polarization; in contrast to strips of high spin orbit material such as Pt which only possess the spin Hall effect and are only sensitive to an in-plane spin polarization of the spin accumulation. We show that nonlocal magnon transport is able to measure the spin-flop transition in the absence of other spurious effects. Our measurements show the detection of magnons generated by the spin Seebeck effect before and after the spin-flop transition where the signal reversal of the magnon spin accumulation agrees with the OOP spin polarization carried by magnon modes before and after the SF transition. 
	\end{abstract}

	\maketitle
	
The recent discovery of long range magnetic ordering in two dimensional magnets \cite{Xing2019,Liu2020a} opens possibilities to study and explore the magnetic structure and dynamics in two dimensional magnets.  Especially antiferromagnetic materials have gained a great interest for information storage and as a medium for spin currents in spintronic devices because they do not possess stray fields, are robust against magnetic perturbations and have ultra-fast magnetic dynamics \cite{Jungwirth2018b,Baltz2018}. Antiferromagnets possess a variety of spintextures, for example uniaxial, easy-plane or noncolinear spintextures, determined by the material specific values of the exchange field and anisotropy field parameters. Additionally, magnetic van der Waals materials have often much stronger intralayer exchange interactions than interlayer exchange interactions giving magnetic van der Waals materials a rich variety in spintextures.

Characterizing and probing magnetic transitions in (quasi) 2 dimensional magnetic van der Waals materials is crucial to understand magnetism at a low dimensional limit; for example by characterizing the spin-flop (SF) transition in uniaxial antiferromagnets (AFM).  When the applied magnetic field $H_{0}$ exceeds the spin-flop field $H_{\text{SF}}$ at the SF transition, the spin configuration changes from (anti) parallel to (almost) perpendicular to the magnetic field. The SF transition has been studied by magnetic measurements \cite{Basnet2021}. This is difficult for thin (small volume) magnetic layers however as an alternative magnons can be used to study the SF transition electrically. At the spin-flop field, the energy for certain magnon modes goes to zero which should result in a strong modification and even sign reversal of the spin polarization of the magnon generated by the spin Seebeck effect. 
\begin{figure}[h!]
	\includegraphics[width=80mm]{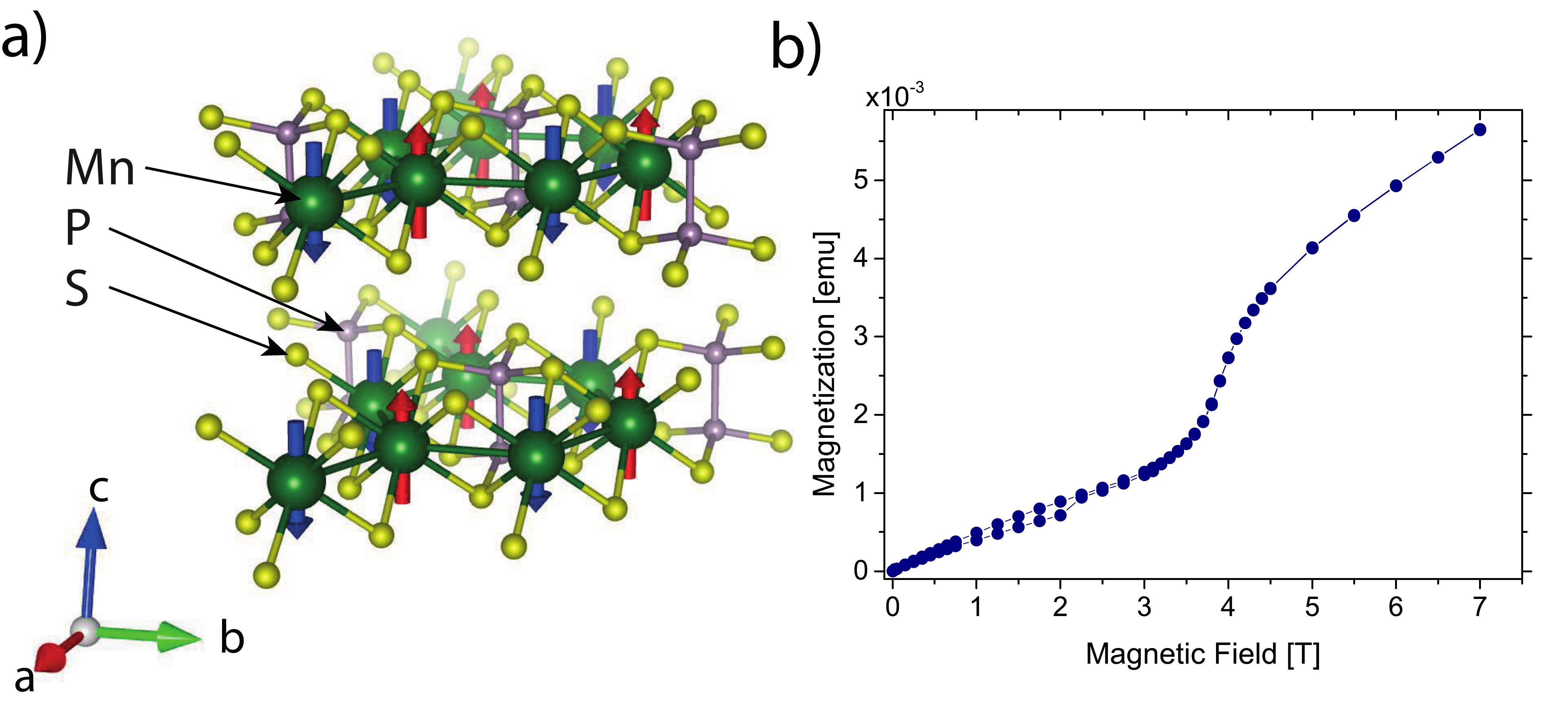}
	\caption{(a) Spin structure of MnPS$_{3}$. The red and blue arrows denote the spin direction in the absence of a magnetic field. (b) Magnetization measured for an applied out-of-plane magnetic field. The spin-flop transition is observed around 3.7\,T, indicated by the sharp increase of the magnetization.}
	\label{fig:Image1}
\end{figure}
\begin{figure*}[th!]
	\centering
	\includegraphics[width=1\linewidth]{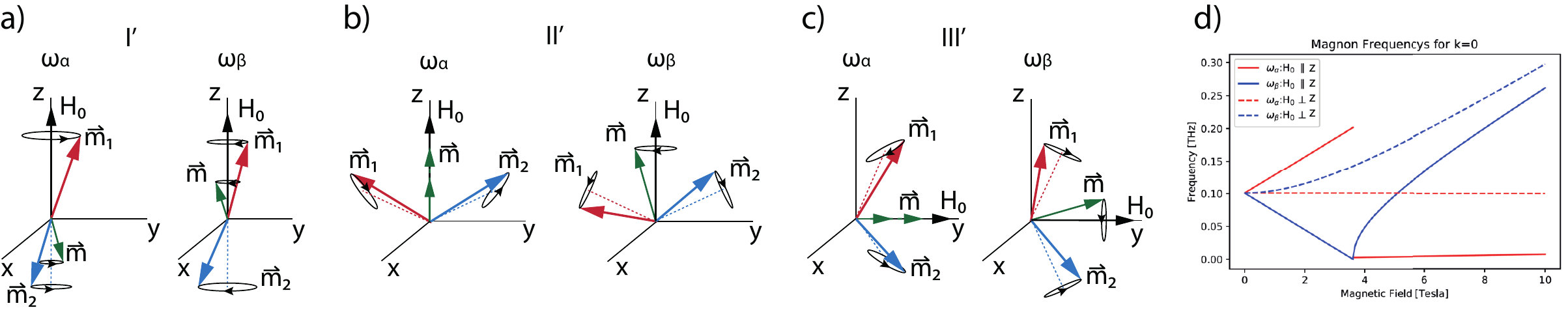}
	\caption{(a), (b) and (c)  The magnon modes I', II' and III' for the corresponding spin configurations I, II and III. (a) I' possesses two circular polarized magnon modes $\omega_{\alpha}$ and $\omega_{\beta}$. (b) After the spin-flop transition, II' possesses a magnon mode $\omega_{\alpha}$ linearly polarized in $\vec{n}$ and $\vec{m}$, and a magnon mode  $\omega_{\beta}$ which is linearly polarized in $\vec{n}$ and circular polarized in $\vec{m}$. (c) III' possesses similar magnon modes as for II' but the order parameters $\vec{n}$ and $\vec{m}$ point in a different direction. (d) k = 0 frequencies for the magnon modes I', II' and III' are plotted. }
	\label{fig:Image3}
\end{figure*}

Magnon spintronics uses magnons to transport angular momentum which is an unique tool to investigate magnetic dynamics in magnetic materials because it can characterize magnetic van der Waals materials down to a monolayer using a heterostructure of a heavy (ferromagnetic) metal and the magnetic layer. Magnon transport has been extensively studied in 3 dimensional magnets via spin pumping \cite{Costache2006b, Kajiwara2010, Chumak2012}, spin Seebeck effect \cite{Uchida2008, Uchida2010a, Wu2016a} and electrical injection and detection \cite{Cornelissen2015, Goennenwein2015}. Nonlocal magnon transport has been observed in ferrimagnets \cite{Cornelissen2015,Shan2016} and antiferromagnets \cite{Lebrun2018b,Hoogeboom2020,Yuan2018,Wimmer2020,Xing2019a}. 
It has been shown that the SF transition in Cr$_{2}$O$_{3}$ (3D), for which the spins lie in-plane before and after the SF transition, can be probed locally via the spin Seebeck effect using a Pt \cite{Li2020,Reitz2020} and Py \cite{Rodriguez2022} contacts. The spin Hall magnetoresistance detected the SF transition in Pt or Pd in contact with the van der Waals AFM CrPS$_{4}$. This was detected locally and therefore SMR only probes the magnetic properties in the first layer(s) of the AFM in contact to the heavy metal \cite{Wu}. Detecting the SF transition via thermally generated magnons in a nonlocal geometry in van der Waals magnets has not been investigated. This has the benefit of studying the SF transition in a magnetic material outside the proximity of a heavy metal. Moreover, as we will show, no spurious (thermal) effects are present in the detector strips except for an anomalous Nernst effect which can be subtracted in a straightforward way.

\begin{figure*}[th!]
	\centering
	\includegraphics[width=1\linewidth]{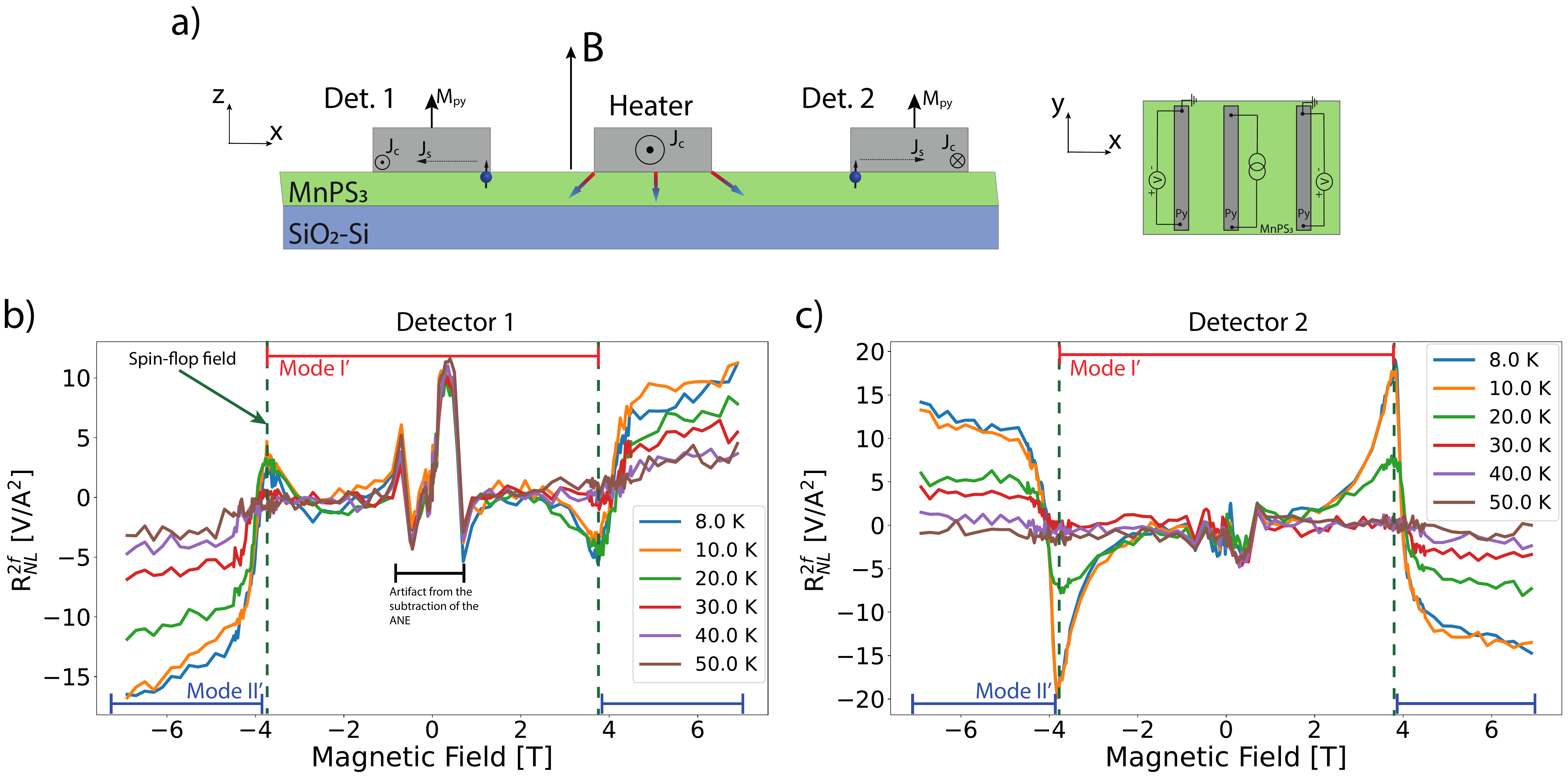}
	\caption{(a) Illustration of the device geometry. A Py heater and detector strips on MnPS$_{3}$ are used and a magnetic field is swept perpendicular to the crystallographic a-b plane from -7\,T to 7\,T. Second-harmonic resistance for (b) detector 1 and (c) detector 2 are shown. A current of 100 $\mu A$ is applied at the heater strip generating a temperature gradient in MnPS$_{3}$. The measurements are performed at 8 Kelvin and the seperation between the heater and the strips are 1\,$\mu$m and 2\,$\mu$m. The structure in the data between -1 and 1 T is due to an artifact of subtracting the ANE contribution from the data \cite{Sup1}.}
	\label{fig:R2w_OOP}
\end{figure*}

In this work, we detect the SF transition in antiferromagnetic transition metal trichalcogenide MnPS$_{3}$ using thermally generated magnons which we detect nonlocally using Pt and Py contacts. A heater generates a temperature gradient in MnPS$_{3}$ which generates magnons via the spin Seebeck effect which are detected at a Pt or Py detector (Fig. \ref{fig:R2w_OOP}). This is very clean way of measuring due to the lack of other spurious effects in the detector strips and in the absence of a temperature gradient across the interface of the detector strips \cite{Sup1}. The magnetic easy axis in MnPS$_{3}$ is out-of-plane (OOP), i.e. perpendicular to the $a-b$ plane, and therefore the generated magnons carry spins with an OOP polarization. These spins cannot be detected via the regular inverse spin Hall effect (ISHE) which only generates a charge current for a spin current with an in-plane spin polarization. However the inverse anomalous Hall effect (IASHE) is able to detect a spin current with an OOP spin polarization, for example using Py contacts \cite{Das2017a,Das2018a}. We show that the Py contacts can detect the spin-flop transition in MnPS$_{3}$ via magnons which carry spins with an OOP spin polarization via the IASHE in Py. We detect that the polarization of the spins carried by the magnon changes sign when crossing the SF transition and that the detected signal is maximum when the energy of the relevant magnon modes go to zero.

	Antiferromagnets are characterized by two order parameters, the Néel vector  $\vec{n} = \vec{m_{1}} - \vec{m_{2}}$ and the net magnetization $\vec{m}= \vec{m_{1}} + \vec{m_{2}}$.  An easy-axis antiferromagnet undergoes a SF transition when the applied magnetic field strength along the easy axis exceeds the spin-flop field $H_{\text{SF}} = \sqrt{2H_{\text{A}}H_{\text{E}} - H_{\text{A}}^{2}} $, where $H_{\text{A}}$ is the anisotropy field and $H_{\text{E}}$ is the exchange field strength of the antiferromagnetic material. After the spin-flop transition, the spins cant towards the applied magnetic field direction.

Dynamically, antiferromagnets possess a variety of magnon modes depending on the state of the antiferromagnet, presented in Fig. \ref{fig:Image3}(a), (b) and (c). Below the SF transition, the degeneracy of the two magnon modes is lifted by the Zeeman splitting when a magnetic field is applied and therefore one magnon mode, $\omega_{\beta}$ decreases and magnon mode $\omega_{\alpha}$ increases in energy, as shown in Fig. \ref{fig:Image3} (d) for the $k=0$ magnon modes. When exciting the magnon modes at a finite temperature, the lower energy mode, $\omega_{\beta}$ is populated more which spin is oriented \textit{along} the magnetic field direction. Above the SF, the magnons, $\omega_{\beta}$, carry spins with a polarizion in the z direction \textit{opposite} to the magnetization order parameter $\vec{m}$ and therefore the spin polarization in the z direction of the magnons are \textit{opposite} to the applied magnetic field direction. Consequently, the spin polarization direction of the magnons changes from parallel to anti parallel to the applied magnetic field direction crossing the SF transition. 

	In a magnetic insulator a magnon spin current can be generated due to a temperature gradient in a magnetic material via the spin Seebeck effect (SSE) which can be expressed as \cite{Rezende2019a}:
	\begin{equation}
		J^{z}_{S} = S^{z}_{S}\,\nabla T
	\end{equation}
	
	At zero magnetic field the magnon modes I' are degenerate and therefore under the influence of a temperature gradient both modes are populated equally. Both modes carrier opposite angular momentum and therefore no net spin current is present. An imbalance in population between the modes is present at a finite magnetic field resulting in a finite net spin current. The spin Seebeck coefficient $S^{z}_{S}$ depends on the difference in occupation of the different magnon modes \cite{Sup2}. At the SF transition, mode $\omega_{\beta}$ reaches zero and therefore the largest $S^{z}_{S}$ is expected. 

	The total detected voltage is given by \cite{Rezende2014a}:	
	\begin{equation}\label{Volt}
		V_{\text{NL}} = G S^{z}_{S} 
	\end{equation}
	where $G \propto  R_{N}w\lambda_{N} \frac{2e}{\hbar} \theta_{ASH} \tanh(\frac{t_{N}}{2\lambda_{N}}) C g^{\uparrow\downarrow}\vec{\nabla}T$ with  $R_{N}$ is the resistance, $t_{N}$ the thickness, $w$ the width, $\theta_{ASH}$ the anomalous spin Hall angle and $\lambda_{N}$ is the spin relaxation length of the detector strip.  $g^{\uparrow\downarrow}$ is the effective spin mixing conductance and $C$ contains the thickness, magnon diffusion length, geometry of the nonlocal device and other material parameters.

	The transition metal trichalcogenide crystals were bought commercially from the company HQ Graphene \cite{HQ}. The crystals were magnetically characterized using a magnetic property measurement system (MPMS) to extract the Néel temperature and the magnetization behavior. Fig. \ref{fig:Image1}(b) shows the magnetization measurement for an out-of-plane magnetic field. MnPS$_{3}$ possesses an uniaxial anistropy perpendicular to the a-b plane. This is generated due to a competition between a dipolar interaction which prefers the spins to lie perpendicular to the a-b plane, and a single ion anisotropy which prefers the spins to lie in the a-b plane \cite{Okuda1986_2, wildes2006a}. The anisotropy field in the OOP direction due to dipolar interaction is stronger and therefore the easy axis is perpendicular to the a-b plane. The spin-flop transition is around 3.7\,T, as shown in Fig. \ref{fig:Image1}(b), which shows a similar trend as in \cite{Basnet2021}. The magnetic susceptibility shows a sharp decrease below 80K which is the Neél temperature of MnPS$_{3}$ \cite{Sup3}.
	
	The crystals were mechanically exfoliated onto a silicon silicon oxide substrate in a nitrogen atmosphere to prevent any oxidation.  Two types of devices were fabricated, one with Py detector strips and one with Pt detector strips. The Py and Pt strips were sputtered on the flakes and were 7~nm thick and 200~nm wide. The centre-to-centre distance between the injector and detector electrodes varies for the different devices between $0.6~\mu\text{m}$ up to $5~\mu\text{m}$. The detector strips were connected to Ti (5\ nm)-Au(50\ nm) leads which were evaporated on top \cite{Sup4}.

	In order to generate magnons thermally, a charge current was sent through the Py and Pt heater strips to generate a temperature gradient due to Joule heating in the strip. The thermally generated magnons diffuse in the magnetic material and were detected nonlocally via spin-flip scatting with electrons in Pt or Py, generating a spin accumulation in the Pt or Py detector strip. 
	
\begin{figure*}[th!]
	\hspace*{-8mm}
	\centering
	\includegraphics[width=150mm]{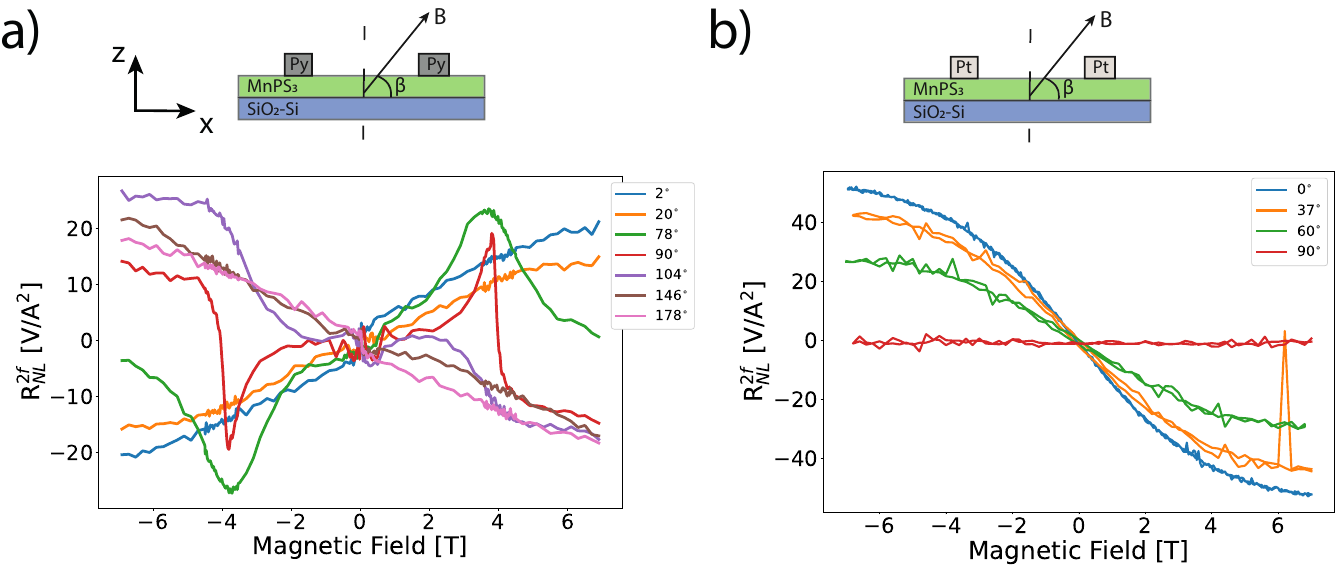}
	\caption{$R_\text{NL}^\text{2f}$ response for (a) Py and (b) Pt heaters and detector strips for a 7T magnetic field applied at various angles of $\beta$ at 8K. Magnon modes are excited which carry spins polarized in the OOP and/or IP direction depending on the magnetic field direction. The magnons arrive at the detector strips where the Py detectors can detect spins polarized in the IP and OOP direction whereas Pt detectors are only sensitive to IP polarized spins. Comparing Py and Pt detectors, it is clear that at all angles the Pt detector strips do not detect the SF transition. In contrast, the OOP polarized spins are a significant contribution to the spin polarization direction of the magnon modes.}
	\label{fig:Image4}
\end{figure*}

	In Pt a charge current was generated via the inverse spin Hall effect (ISHE), which converts a spin current into a charge current. In a ferromagnetic heavy metal a charge current was generated via the ISHE and the inverse anomalous spin Hall effect (IASHE) \cite{Das2017a}. The magnitude and directions of the spin current detected by the IASHE depends on the product $\vec{J}_{\text{s}} \times \vec{M}_{\text{Py}}$  for spins polarized in the direction of $\vec{M}_{\text{Py}}$. Therefore a Py detector strip can detect an out-of-plane spin accumulation which is controlled via the magnetization direction of Py \cite{Das2018a, Sup5}. 
	
	An alternating current ($I$) was sourced through the Pt and Py injector. The second harmonic (2f) responses of the nonlocal voltage ($V$) was measured across the Pt and Py detectors via a lockin amplifier. The nonlocal resistance is defined as $R_\text{NL}^\text{2f}=V^\text{2f}/I^2$ \cite{Cornelissen2015}.

Figure \ref{fig:R2w_OOP} shows $R_\text{NL}^\text{2f}$ for magnetic field sweeps with a magnetic field applied out-of-plane for a device with a Py heater and detector strips at various temperatures. An anomalous Nernst contribution due to a temperature gradient in the x-axis direction in the Py strips is subtracted from the measured signal \cite{Sup1}.  $R_\text{NL}^\text{2f}$ shows a increase of the signal with magnetic field with a maximum at the SF field which corresponds to the frequency of magnon mode I' $\omega_{\beta}$ reaching zero as shown in Fig. \ref{fig:Image3} (d). At the SF field an abrupt sign change is observed in the data which corresponds to the transition of generated magnon modes from Fig. \ref{fig:Image3} (a) to (b). This transition results from a sign change of the spin polarization of the magnon modes, similar to observation in Fig. \ref{fig:R2w_OOP} (b) and (c). 

The direction of the generated charge current in detector 1 and 2 is opposite because the direction of the spin current in detector 1 and 2 is in opposite direction, as shown in Fig. \ref{fig:R2w_OOP}. The charge current generated in the Py strip due to the IASHE depends on $\vec{J}_{c} \propto \theta_{\text{ASH}}\vec{J}_{\text{s}} \times \vec{M}_{\text{Py}}$. For an OOP applied magnetic field,  $\vec{J}_{c}$ is zero for a $\vec{J}_{\text{s}}$ in the z direction and non zero for a $\vec{J}_{\text{s}}$ in the x direction \cite{Sup5}. Therefore the direction of $\vec{J}_{c}$ is opposite because the direction of $\vec{J}_{\text{s}}$ in detector 1 is in -$\hat{x}$ and in detector 2 in the $\hat{x}$ direction.
 
The applied magnetic field direction determines which magnon modes are excited in the AFM, for example for an IP magnetic field magnon mode III' is excited and for an OOP magnetic field I' is excited. Therefore, the polarization direction of the spins carried by the magnon modes changes when changing the direction of the applied magnetic field. Next to that, the IASHE can detect an IP and OOP polarization of the spin current in Py depending on the orientation $\beta$ of $\vec{M}_{\text{Py}}$ and the direction of the spin current in Py. Therefore the $R_\text{NL}^\text{2f}$ response for various directions of the applied magnetic field, as shown in Fig. \ref{fig:Image4}, is a combination of the detection spins polarized in te IP and OOP direction. For $\beta = 0^{\circ}$ the detected spin polarization direction is purely IP (x direction) and for $\beta = 90^{\circ}$ purely OOP (z direction).

A device consisting of Pt heater and detectors is only sensitive to an IP spin polarization direction because it only possesses the ISHE. Therefore $R_\text{NL}^\text{2f}$ is maximum for $\beta = 0^{\circ}$, as shown in Fig. \ref{fig:Image4} (b). Spins with an OOP spin polarization direction are not detected and therefore the SF transition is not detected via the Pt detector strips. Comparing the result of Fig. \ref{fig:Image4} (a) and (b) show that at angles between $\beta = 0^{\circ} - 90^{\circ}$ magnons with a OOP spin polarization are a significant contribution of the generated magnons due to the SSE and that the IP polarized spins carried by the magnons seems to be completely insensitive to the SF transition.

\begin{figure}[th!]
	\hspace*{0mm}
	\centering
	\includegraphics[width=\linewidth]{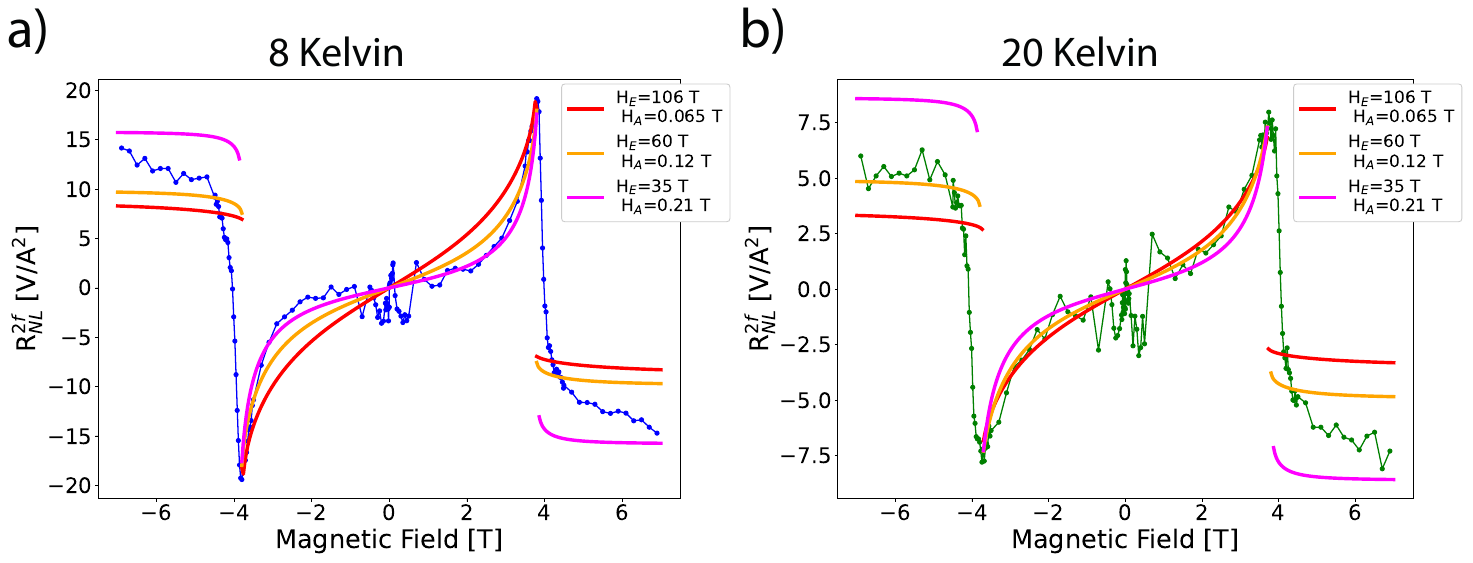}
	\caption{Measurement shown in Fig. \ref{fig:R2w_OOP} (b) for 8 and 20 K fitted for the magnetic field values below the SF for various values of H$_{\text{E}}$ and H$_{\text{A}}$, where H$_{\text{E}}$ = 106 T and H$_{\text{A}}$ = 0.065 T are the values reported in literature \cite{Okuda1986_2}. We have used $g^{\uparrow\downarrow}_{\text{BSF}}$/$g^{\uparrow\downarrow}_{\text{ASF}}$ = 10 to calculate the response after the SF in order to have the correct order of magnitude. The ratio $g^{\uparrow\downarrow}_{\text{BSF}}$/$g^{\uparrow\downarrow}_{\text{ASF}}$ is opposite to the ratio found for local measurements on Cr$_{2}$O$_{3}$-Pt \cite{Reitz2020} where $g^{\uparrow\downarrow}_{\text{ASF}}$ needs to be larger in order to fit the data. Next to that, the values for $C$ vary between the different fits, for more details see \cite{Sup2}. }
	\label{fig:SSE_Fit_BSF_ASF}
\end{figure}

The expected signal due to the SSE before and after the spin-flop transition, SSE$_{\text{OOP-BSF}}$ and SSE$_{\text{OOP-ASF}}$, can be calculated using Eq. (\ref{Volt}), where the calculation of S is presented in the supplementary materials \cite{Sup2}. The expected signal due to the SSE is plotted as solid lines in Fig. \ref{fig:SSE_Fit_BSF_ASF}. The magnitude of the expected signal depends on the factor G, Eq. \ref{Volt}, whereas the shape of the expected signal is due to the values of H$_{\text{E}}$ and H$_{\text{A}}$. 
The calculated second-harmonic response has been plotted for various values of H$_{\text{E}}$ and H$_{\text{A}}$, where the ratio of every combination of H$_{\text{E}}$ and H$_{\text{A}}$ is kept such that the spin-flop field, H$_{\text{SF}}$, is the same. Figure \ref{fig:SSE_Fit_BSF_ASF} suggests that at 8 K H$_{\text{A}}$ is increased and H$_{\text{E}}$ is reduced, whereas at 20 K the values of H$_{\text{A}}$ and H$_{\text{E}}$ are comparable to the literature values \cite{Okuda1986_2, Sup2}.

Possible explanations for the change in values of H$_{\text{A}}$ and H$_{\text{E}}$ at lower temperatures could be that the single ion contribution to the anisotropy is reduced and therefore increasing the anisotropy strength in the OOP direction due to the dipolar interaction. On the other hand, this would not explain why the exchange interaction strength would reduce. The magnon relaxation rate is kept constant in this calculation which could be temperature and magnetic field strength dependent and t
Another explanation could be that the magnon relaxation rate is temperature and magnetic field strength dependent which is kept constant in this calculation. This should be evaluated by an extensive analyses of the the temperature and field dependence of the relaxation rate which is beyond the scope of this paper.

In conclusion we detected the spin-flop transition in MnPS$_{3}$ using nonlocal magnon transport generating the magnons thermally and detecting them via the inverse anomalous spin Hall effect using Py detector strips. The measurements are compared to nonlocal magnon transport using Pt detector strips which are not capable of detecting an out-of-plane spin accumulation, showing the unique power of the IASHE. We detect that the magnons change sign when crossing the spin-flop transition as expected from the magnon modes generated across the SF transition. Next to that, we observe a change in the exchange and anistropy fields of MnPS$_{3}$ reaching temperatures below 10 K when we compare the experimental results to the calculated SSE coefficient.  \\

\begin{acknowledgments}
	We want to thank Dr. G. R. Hoogeboom and Dr. J. Peiro  for  the  useful  scientific  discussions and feedback on the interpretation of the data. We acknowledge the technical support from J.\ G.\ Holstein, H.\ Adema T.\ Schouten and H. de Vries. We acknowledge the financial support of the Zernike Institute for Advanced Materials. This project is also financed by the 2016 NWO Spinoza prize awarded to Prof.\ B.\ J.\ van Wees.
\end{acknowledgments}

\bibliographystyle{apsrev4-2}
%
\end{document}